\documentstyle[12pt]{article}

\newcommand{\EQ}{\begin{equation}}
\newcommand{\EN}{\end{equation}}
\def\aprle{\buildrel < \over {_{\sim}}}
\def\aprge{\buildrel > \over {_{\sim}}}
\begin{document}
\topmargin 0pt
\oddsidemargin=-0.4truecm
\evensidemargin=-0.4truecm
\renewcommand{\thefootnote}{\fnsymbol{footnote}}
\newpage
\setcounter{page}{0}
\begin{titlepage}
\begin{flushright}
FTUV/94/64,
IFIC/94/67\\
SISSA 169/94/A-EP \\
\end{flushright}
\begin{center}
{\large SOLAR NEUTRINO DATA, NEUTRINO MAGNETIC MOMENTS\\
AND FLAVOR MIXING}\\
\vspace{0.5cm}
{\large E.Kh. Akhmedov}
\footnote{On leave from NRC ``Kurchatov Institute'', Moscow 123182, Russia.
Present address: Scuola Internazionale Superiore di Studi Avanzati, Via Beirut
2--4, I-34014 Trieste, Italy} \\
{\em Instituto de Fisica Corpuscular (IFIC--CSIC)\\
Departamento de Fisica Teorica, Universitat de Valencia\\
Dr. Moliner 50, 46100 Burjassot (Valencia), Spain}\\
\vspace{0.5cm}
{\large A. Lanza and~S.T. Petcov}\footnote{Istituto Nazionale di Fisica
Nucleare, Sezione di Trieste, Italy}
\footnote{Institute of Nuclear Research and Nuclear Energy, Bulgarian
Academy of Sciences, BG-1784
Sofia, Bulgaria}\\
{\em Scuola Internazionale Superiore di Studi Avanzati\\
Via Beirut 2--4, I-34014 Trieste, Italy} \\
\end{center}
\begin{abstract}
The results of all currently operating solar neutrino experiments are
analyzed in the framework of the resonant neutrino spin--flavor precession
scenario including the effects of neutrino mixing.
Nine different profiles of the solar magnetic field are used
in the calculations. It is shown that the available experimental data
can be accounted for within the considered scenario. The Ga--Ge data lead
to an upper limit on the neutrino mixing angle: $\sin 2\theta_0 \aprle 0.25$.
One can discriminate between small mixing angle ($\sin 2\theta_0 \aprle 0.1$)
and moderate mixing angle solutions by studying the solar $\bar{\nu}_{e}$
flux which is predicted to be sizeable for moderate mixing angles. The expected
signals due to $\bar{\nu}_{e}$
in the SNO, Super--Kamiokande and Borexino experiments are
calculated and found to be detectable for $\sin 2\theta_0 \aprge 0.1$.
\end{abstract}
\end{titlepage}
\renewcommand{\thefootnote}{\arabic{footnote}}
\setcounter{footnote}{0}
\newpage
\section{Introduction}
The solar neutrino problem, i.e. the discrepancy between the solar neutrino
observations \cite{Davis,Lande,KamVen94,NewGALLEX,SAGE} and the solar model
predictions \cite{B,TCh,Proff,Cast,DS}
remains one of the major unresolved puzzles of modern particle physics and
astrophysics. Although at present an astrophysical solution of the problem is
not completely ruled out, it is rather unlikely to be the true cause
of the discrepancy provided the results of the Cl--Ar [1,2], Kamiokande [3],
and the Ga--Ge [4,5] experiments are correct \cite{Discu,Cast}.

There are several possible neutrino--physics solutions of the solar
neutrino problem, the most popular one being resonant neutrino transitions
in the matter of the sun (the MSW effect \cite{MSW}). In this paper we
concentrate, however, on another type of solutions related to the hypothesis
of existence
of relatively large magnetic or transition magnetic moments of neutrinos.
In this case neutrino spin precession \cite{C,VVO} or spin--flavor
precession \cite{VVO,Akhm1,LM} can occur in the magnetic field of the sun,
converting a fraction of the solar $\nu_{eL}$'s into $\nu_{eR}$ or into
$\nu_{\mu R}$, $\nu_{\tau R}$, $\bar{\nu}_{\mu R}$ or $\bar{\nu}_{\tau R}$.
Although $\bar{\nu}_{\mu R}$ and $\bar{\nu}_{\tau R}$ are not sterile, they
cannot be observed in the Cl--Ar (Homestake), and the Ga--Ge (SAGE and
GALLEX) experiments, and can only be detected with a small cross section in
the Kamiokande experiment. Spin--flavor precession of neutrinos can be
resonantly enhanced in the matter of the sun \cite{Akhm1,LM}, in direct
analogy with the MSW effect.

Resonant spin--flavor precession (RSFP) of neutrinos can account for both the
deficiency of solar neutrinos and the time variations of the solar neutrino
flux in anticorrelation with the solar activity, for which there are some
indications in the Homestake data \cite{tvar,BSSS}. Such an anticorrelation
can be related to the fact that the magnetic field of the sun is strongest
in the periods of active sun.

The Homestake data is fitted better using the hypothesis of a time--dependent
signal than that of a constant one: analyses performed exploiting different
statistical methods produced fairly large values of the coefficient of
correlation between the
data and sunspot number \footnote{ As is well known, the sunspot number
provides a quantitative measure of the solar activity.}
\cite{tvar,BSSS}. These analyses were completed,
however, before 1990 and so did not take into account the data available from
the more recent Homestake runs 109--126. The data from the runs 109--126
do not exhibit a tendency to vary with time, similarly to the data from the
runs 19--59. A recent
analysis of Stanev \cite {St}, which updated the one of ref. \cite{BSSS},
included the results from the runs 109--126. It showed that this leads to
the correlation
coefficient being decreased by an order of magnitude as compared to the
previously obtained one, but the correlation probability still remains large:
the confidence level of the correlation with the sunspot number $s$ is 0.96
instead of 0.996, and that of the correlation with $s|z|$, where $z$ is the
latitude of the line of sight, is 0.99 instead of 0.9993.  The correlation
with the 22-yr cycle is even better than the correlation with the 11-yr
one. Therefore, the possibility that the solar neutrino flux
anticorrelates with solar activity still persists and deserves a further
study.

At the same time, the Kamiokande group did not observe time variations
of the solar neutrino signal in their experiment, which allowed them to put
an upper limit on the possible magnitude of the effect, $\Delta Q/Q < 30\%$
at 90\% c.l. \cite{KamVen94}. Thus, the question naturally arises as to
whether one can reconcile a relatively strong time variation of the signal
in the Homestake experiment with a small (or no) time variation of the
Kamiokande event rate.

Recently, it has been shown \cite{ALP1} (see also
\cite{Kr,MN2}) that the RSFP scenario is capable of accounting for
all the existing solar neutrino data, including their time structure or lack
of such a structure. In particular, it can naturally reconcile sizeable time
variations of the signal in the Homestake experiment with small time
variations allowed by the Kamiokande data. The key points here are that
\cite{Akhm2,BMR,OS,ALP1} i) the two experiments are sensitive to slightly
different parts of the solar neutrino spectrum, and  ii) the RSFP can convert
left--handed $\nu_{e}$ into right--handed
$\bar{\nu}_{\mu}$ (or $\bar{\nu}_{\tau}$) which are sterile for the Homestake
experiment, but do contribute to the event rate in the Kamiokande experiment
through their neutral--current interaction with electrons.

The RSFP mechanism can also explain mild suppression of the signal
in the Ga--Ge experiments. Most of the GALLEX and SAGE data have been
taken during the period of high solar activity. Therefore one could expect a
strong suppression of the signals in these experiments, which has not been
observed. This disfavors the ordinary spin precession scenario since it is
neutrino-energy independent and so predicts universal suppression
and time variation of the signals in all solar neutrino experiments.
On the contrary, the RSFP is strongly neutrino-energy
dependent which naturally leads to different degrees of suppression and
time variation of the event rates in different experiments. In particular,
the $pp$ neutrinos
which are expected to give the major contribution to the signal in the
Ga--Ge experiments, have low energies and so should encounter the RSFP
resonance at higher densities than the $^8$B and $^7$Be neutrinos
(the resonant density is inversely proportional to neutrino energy).
We know that the magnetic field does exist and may be quite strong in the
convective zone of the sun ($0.7R_\odot\aprle r \aprle R_\odot$).
If the $^8$B and $^7$Be neutrinos experience the RSFP
conversion in the convective zone (which is needed to account for the
Homestake and Kamiokande data), the $pp$ neutrinos will encounter the RSFP
resonance somewhere in the radiation zone or in the core of the sun.
However, it is not clear
if a sufficiently strong magnetic field can exist deep in the sun, i.e. in
the radiation zone or in the solar core. If the inner magnetic field of the
sun is week, the RSFP will not be efficient there and the $pp$ neutrinos will
leave the sun intact, in accordance with the observations of GALLEX and SAGE.
One can turn the argument around and ask the following question: What is the
maximal allowed inner magnetic field which is not in conflict with the
Ga--Ge data? The answer turns
out to be $(B_i)_{max}\approx 3\times 10^6$ G assuming neutrino
transition magnetic moment $\mu=10^{-11}\mu_B$ \cite{ALP1}.

Recently, we have analyzed all the available solar neutrino data in the
framework of the RSFP disregarding the neutrino flavor mixing \cite{ALP1}. In
the present paper we extend our previous study to include neutrino mixing and
oscillation effects. Our motivation for that was as follows:

(1) RSFP requires non-vanishing flavor-off-diagonal neutrino magnetic
moments, i.e., implies lepton flavor non-conservation. In general,
one should  therefore
consider the RSFP and neutrino oscillations (including the MSW effect)
jointly. The results of ref. \cite{ALP1} are only valid in the small
neutrino mixing angle limit.

(2) It has been shown in \cite{ALP1} that all the existing solar neutrino data
can be fitted within the RSFP scenario for certain model magnetic field
profiles and certain values of neutrino parameters $\mu$ and $\Delta m^2$.
It would be interesting to see how the neutrino mixing modifies these
results.

(3) In ref. \cite{Akhm3} it has been suggested that the combined action of
the RSFP and MSW effect in the convective zone of the sun can relax by a
factor of 2--3
the lower limit on the product $\mu B_{\bot}$ of the neutrino magnetic moment
and solar magnetic field strength required to account for the data. The main
idea was that the MSW effect can assist the RSFP to cause the time variations
of the neutrino flux by improving the adiabaticity of the RSFP (this can occur
when the RSFP and MSW resonances overlap). It would be interesting to
confront this idea with the new experimental data.

  The combined action of the RSFP and the MSW effect on solar neutrinos has
been considered in a number of papers \cite{LM,Akhm4,BalHL}. However,
the data of the Ga--Ge experiments were not available at that time. In the
present paper we analyze all the existing solar--neutrino data including
those of the gallium detectors. We also give predictions for the
forthcoming solar neutrino experiments paying special attention to the
possibility of detection of $\bar{\nu}_e$'s coming from the sun.

\section{Neutrino propagation in the sun and solar magnetic fields}
We consider transitions of solar neutrinos in the two-flavor
approximation assuming massive neutrinos to be Majorana particles with a
transition magnetic moment $\mu$ and taking into
account the neutrino mixing and matter effects in the sun. The neutrino basis
is taken to be ($\nu_e$, $\bar{\nu}_e$, $\nu_{\mu}$, $\bar{\nu}_{\mu}$).
The transverse (toroidal) magnetic field of the convective zone of the sun
$B_{\bot}$ is assumed to have a fixed direction
\footnote{The case of magnetic field which rotates in the transverse plane
along the neutrino trajectory (``twisting'' magnetic field) has been
considered in \cite{Sm,Kr,APS}. }.
The evolution equation for such a neutrino system was given in \cite{LM}.

Unfortunately, very little is known about the magnetic field of the sun.
Not only its profile is unknown, but even its strength is in fact very
uncertain. Energy balance consideration yield only rather loose bounds
on the solar magnetic fields strength. One is therefore forced to use various
more or less ``plausible'' model magnetic field configurations.
With a very precise data on the flux and spectrum of solar neutrinos and
independent information about the neutrino magnetic moments one could
in principle solve the inverse problem and get an
information about the solar magnetic field.
At the moment the only way to proceed is to compute the signals in various
solar--neutrino detectors and to compare them with the data. This approach
allows one even now to discard certain magnetic field configurations which
fail to simultaneously account for all the existing data.

We have calculated the neutrino signals in the Homestake, Ga--Ge and
Kamiokande experiments using nine different model magnetic field profiles in
the convective zone, assuming that the field is absent in the radiation zone
and in the core of the sun. It was also assumed that the magnetic field
strength varies in time in direct correlation with solar activity. The
minimum and maximum magnetic field strengths were fixed to reproduce the
apparent time variations of the Homestake signal (see \cite{ALP1} for more
details). In most of the cases we have considered it was assumed that the
solar magnetic field vanishes at
the surface of the sun. The only exception was the magnetic field
configuration COSH (see below) in which the magnetic field was assumed to
die off exponentially. We therefore integrated the evolution equation
numerically taking into account neutrino flavor oscillations
($\nu_{eL}\leftrightarrow \nu_{\mu L}$ and $\bar{\nu}_{eR}
\leftrightarrow \bar{\nu}_{\mu R}$) along the whole neutrino path between the
core of the sun and the earth and spin--flavor transitions
($\nu_{eL}\leftrightarrow \bar{\nu}_{\mu R}$ and $\bar{\nu}_{eR}
\leftrightarrow \nu_{\mu R}$) only in the convective zone of the sun,
$0.7R_\odot\aprle r\aprle R_\odot$; for for the COSH magnetic field
configuration the integration of the evolution equations describing the RSFP
was extended until $r = 1.3R_\odot$.

The magnetic field configurations used were:
\EQ
B_{\bot}(x)=B_0\left[1-\left( \frac{x-0.7}{0.3}\right)^n \right],
{}~~~0.7\leq x \leq 1,
\EN
where $x\equiv r/R_\odot$ and $n$=2, 6, 8 and 10 (hereafter referred to as
MAG$n$ configurations);
\EQ
B_{\bot}(x)=B_0\left[1-\left( \frac{x-0.9}{0.1}\right)^6 \right],
{}~~~0.8\leq x \leq 1,
\EN
(MAG69 configuration); the same as eq. (2) with an additional horizontal
line $B_{\bot}=B_0/2$ between the point $x=0.7$ and the point where this
horizontal line touches the profile of eq. (2), i.e.
\EQ
B_{\bot}(x)=\left\{\begin{array}{c}
B_0/2, ~~~0.7\leq x \leq 0.811, \\
B_0\left[1-\left( \frac{x-0.9}{0.1}\right)^6 \right],
{}~~~0.811 \leq x \leq 1,
\end{array}\right.
\EN
(MAG69+ configuration);
\EQ
B_{\bot}(x)=\left\{\begin{array}{c}
B_0\frac{x-x_{0}}{x_c-x_{0}},~~~x_{0}\leq x<x_c,\\
B_0-(B_0-B_f)\frac{x-x_{c}}{1-x_{c}},~~~x_{c}\leq x\leq 1
\end{array}\right.
\EN
with $x_{0}=0.7$, $x_c=0.85$ (LIN2) and 0.9 (LIN9), $B_f=0$ and 100 G,
and
\EQ
B_{\bot}(x)=B_0/\cosh[20(x-0.7)], ~~~0.7\leq x \leq 1.3
\EN
(COSH configuration).
\section{Results and discussion}
The criteria for choosing the minimum and maximum allowed detection rates
(which correspond to high and low convective-zone magnetic field strengths
respectively) were the same as in our previous paper \cite{ALP1} except for
the following modifications. For the Homestake experiment, the average signal
for all the reported data (runs 18--124) was taken to be $2.55 \pm 0.25$ SNU
instead of $2.3 \pm 0.3$ SNU, the increase being related to the changes of
the estimated Ar extraction efficiency (6\%) and counting efficiency (3\%)
\cite{Lande}; the maximum and minimum values of the signal and the iso-SNU
curves were also modified to take these changes of the efficiencies into
account. For the detection rate in the Ga--Ge experiments the combined GALLEX
result for the first 30 runs, $79\pm 12$ SNU, was used \cite{NewGALLEX}. The
1$\sigma$ and 2$\sigma$ iso-SNU curves were redefined in accordance with the
reduced error bars. Qualitatively similar results are obtained using the
SAGE data \cite{SAGE}.

The main conclusions of our analysis are summarized below.

(1) For small mixing angles, $\sin 2\theta_0\aprle 0.1$, the results of
our previous study \cite{ALP1} are only slightly modified. The best fit
of all the data is achieved with LIN2 magnetic field configuration.

(2) For moderate mixing angles, $\sin 2\theta_0\aprge 0.2$, the magnetic
field profile LIN2 which proved to give a good fit of the data for vanishing
$\theta_0$, no longer works: it leads to too strong a suppression of the
signal in the gallium experiments since the MSW transitions of
the low--energy $pp$ neutrinos become adiabatic. Reasonable fit can still
be achieved for very large mixing angles, $\sin 2\theta_0\approx 1$, but in
this case a large flux of electron antineutrinos would be produced, in
contradiction with an upper limit derived from the Kamiokande and LSD data
\cite{BFMM,LSD} (see below, point (5)).

(3) Moderate values of $\theta_0$ are allowed for the magnetic field
profiles whose maximum is shifted towards the outer regions of the
convective zone. For such profiles the RSFP would be efficient for lower
values of $\Delta m^2$ (since the resonance would have to take place
at lower densities), for which in turn the MSW transitions of the $pp$
neutrinos will be non-adiabatic. As a consequence, the flux of the $pp$
neutrinos will be essentially unsuppressed. We have tried three such new
magnetic field configurations (LIN9, MAG69 and MAG69+) and they produced good
fits of all the data.

(4) Typical values of the neutrino parameters required to account for the
data are $\Delta m^2 \simeq (10^{-8}$--$10^{-7})$ eV$^2$, $\sin 2\theta_0
\aprle$ 0.2--0.3, depending on the magnetic field configuration; for neutrino
transition magnetic moment $\mu=10^{-11}\mu_B$ the maximum magnetic field in
the solar convective zone should vary in time in the range (15--45) kG.

(5) As have been noticed above (points (2) and (3)), some magnetic field
configurations which give a good fit to the data for vanishing
$\theta_0$, do not do so for not too small mixing angles and, conversely,
some other profiles which failed to reproduce the data for $\theta_0=0$
do give a good fit for moderate $\theta_0$. This is, in fact, a rather
unpleasant situation: whether or not a given magnetic field profile fits
the data depends on the neutrino mixing angle which is unknown. A possible
way out of this ambiguity is to look for a flux of $\bar{\nu}_{eR}$'s coming
from the sun. If neutrinos experience the RSFP in the sun and also have flavor
mixing, a flux of electron antineutrinos can be produced which in
principle can be detected in the SNO, Super--Kamiokande and Borexino
experiments even in the case of moderate neutrino mixing angles
\cite{LM,Akhm4,RBLBPP,BL1}. We therefore calculated the expected
$\bar{\nu}_{eR}$ signals in these experiments for the magnetic field profiles
and the values of neutrino parameters which fit all the available
solar--neutrino data.

\vspace{0.3cm}
For LIN2 and LIN9 magnetic field profiles defined by eq. (4), it did not
make much difference if we took the surface magnetic field to be 0 or 100 G.
The best fit of all the data for $\sin 2\theta_0 \aprge 0.1$ was achieved
with LIN9 and MAG69+ magnetic field configurations; MAG69 profile was also
able to reproduce the experimental results but gave somewhat smaller allowed
parameter ranges. In figs. 1--5 we present the results of the numerical
calculations for the MAG69+ profile. More complete account of our results
will be presented elsewhere.

Figs. 1--4 give the calculated iso-SNU/iso-suppression curves for the
Homestake, Kamio-kande and Ga--Ge experiments on the planes of the
parameters ($B_0,~\Delta m^2$), ($\sin 2\theta_0,~\Delta m^2$) and
($B_0,~\sin 2\theta_0$). From fig. 1 one can see that for
$\sin 2\theta_0=0.2$
the allowed range of $\Delta m^2$ (which is defined by the right shaded
area since it is more restrictive than the left one) is $\Delta m^2 \approx
(1-3)\times 10^{-8}$ eV$^2$. In fact, with increasing mixing angle the
vertical size of the right shaded area decreases and for $\sin 2\theta_0
\aprge 0.25$ this area disappears. This comes about because the detection rate
in the Ga--Ge experiment becomes too high. The same effect can also be seen in
figs. 2, 3 and 4. The allowed minimum and maximum strengths of the
convective-zone magnetic field of the sun can be found from fig. 1 to be in
the ranges 17--23 kG and 30--50 kG respectively (assuming
$\mu=10^{-11}\mu_B$). This is also illustrated by fig. 4.

In figs. 2 and 3 the allowed ranges of the parameters are given by the
lower shaded areas. The upper ones are excluded by the data since for weak
field (fig. 2) and strong field (fig. 3) they do not overlap.

One can see from fig. 4 that the MAG69+ profile does not fit the data for
zero mixing angle (at least, at the $1\sigma$ level). This illustrates
the point (5) of the above brief summary of our results.

It follows from our analysis that the mixing angles larger than $\sin
2\theta_0 \simeq 0.25$ are excluded since they would lead to too high a
detection rate in the Ga--Ge experiments. This limits the possibility of
having oscillations--assisted RSFP suggested in \cite{Akhm3}. The gain in
the product $\mu B_{\bot}$ which is required for the RSFP to be efficient
turns out to be rather modest: a factor 1.5--2. This can be seen from the
fact that for $\theta_0=0$ and the value of $\mu$ fixed at $10^{-11}\mu_B$
the allowed range of $B_{\bot}$ was 30--50 kG \cite{ALP1} whereas for
$\sin 2\theta_0 = 0.25$ and MAG69+ magnetic field configuration it can be as
low as 17--30 kG.

As we have already mentioned, the combined action of the RSFP and neutrino
oscillations can produce an observable flux of $\bar{\nu}_{eR}$'s from the
sun. We have calculated the fluxes and the corresponding signals in
SNO, Super--Kamiokande and Borexino detectors for various magnetic field
configurations. Fig. 5 shows the energy dependence of the $\nu_{eL}
\rightarrow \bar{\nu}_{eR}$ transition probabilities $P_2$ and $P_{2}^{NVO}$.
The latter takes into account the $\bar{\nu}_{eR}$ production only inside the
sun, i.e. disregards the $\bar{\nu}_{\mu R}\leftrightarrow \bar{\nu}_{eR}$
oscillations in vacuum; it is given just
for comparison. One can see that the main source of the $\bar{\nu}_{eR}$'s
is vacuum oscillations of $\bar{\nu}_{\mu R}$'s (which are
produced in the sun by the RSFP mechanism) into $\bar{\nu}_{eR}$ 's.
This, however, is only true if the transverse magnetic field of the sun has
a fixed direction along the neutrino path. The $\bar{\nu}_{eR}$ flux can be
significantly enhanced in twisting magnetic fields \cite{APS,BL2},
and the $\bar{\nu}_{eR}$ production inside the sun can become more important
than their production due to the vacuum oscillations.
In fact, in this case one can have a detectable $\bar{\nu}_{eR}$ flux even if
the neutrino magnetic moment is too small or the solar magnetic field is too
weak to account for the solar neutrino problem \cite{BL2,AS}.

In fig. 5 also shown are the spectrum of electron antineutrinos as well as
that of the initially produced ${}^8$B neutrinos. Both spectra are normalized
to unit integral, the relative normalization factor being 36.5.
This means that the total flux of electron antineutrinos for maximum
allowed magnetic field is less than $3\%$ of the boron neutrino flux.
Electron antineutrinos from the sun could have been detected by the
Kamiokande and LSD groups
through the $\bar{\nu}_{eR}$--proton capture reaction. Such  antineutrinos
have not been seen which gave an upper limit on the total flux as well as
on the spectrum of solar $\bar{\nu}_{eR}$'s. The integral over the predicted
$\bar{\nu}_{eR}$ spectrum taking into account the instrumental energy
threshold of Kamiokande experiment $E_{th}=7.5$ MeV and the threshold of
the $\bar{\nu}_{eR}$--proton capture reaction turns out to be $1.46\%$ of
the corresponding boron $\nu_{eL}$ flux. This is significantly lower than
the upper bound of $\sim(5-7)\%$ derived from the Kamiokande and LSD data in
\cite{BFMM,LSD}. For the minimum allowed magnetic field strength,
$B_0 \approx$ 20 kG, the $\bar{\nu}_{eR}$ flux is smaller by about a factor
of two. We have also compared the energy spectrum of
$\bar{\nu}_{eR}$ with the Kamiokande limit \cite{Suzuki} and found that
this limit was never violated for the mixing angles $\theta_0$ which fit the
experimental data. This means that at the moment the Ga--Ge results give
more stringent upper bound on $\theta_0$ than the non-observation of solar
$\bar{\nu}_{eR}$'s by the Kamiokande and LSD collaborations.

The forthcoming solar neutrino experiments, such as SNO, Super--Kamiokande
and Borexino, are expected to have much higher sensitivity to
$\bar{\nu}_{eR}$'s than the Kamiokande experiment. We have calculated
expected signals of solar $\bar{\nu}_{eR}$'s in these detectors.
For the magnetic field configuration MAG69+, the event rates for
$\Delta m^2 = 10^{-8}$ eV$^2$ and several values of $\sin 2\theta_0$ are given
in table 1. We have also included the results for the mixing angle
$\sin 2\theta_0=0.32$ which fits the data within $2\sigma$ errors.
The calculated event rates for LIN9 magnetic field
configuration are similar to those for MAG69+.

\newpage
\noindent
\vspace*{-.3truecm}
{\footnotesize Table 1. Predicted numbers of events per year due
to the solar $\bar{\nu}_{eR}$'s in SNO, Super--Kamiokande and
\vspace*{-.3truecm}
Borexino experiments for MAG69+ magnetic field configuration,
$\Delta m^2 = 10^{-8}$ eV$^2$, $B_0=45$ kG. The
\vspace*{-.3truecm}
(instrumental) $e^+$ energy threshold for SNO and Super--Kamiokande is
5 MeV; for Borexino the expected
\vspace*{-.3truecm}
signal for $E_{\bar{\nu}_e}\geq 5$ MeV is given. For Super--Kamiokande the
same $e^+$ detection efficiency as in Kamiokande
is assumed, for SNO and Borexino the detection efficiencies were taken to
be 1.}

\vspace{0.2cm}
\begin{tabular}{|c||c|c|c||c|c|c||} \hline
\multicolumn{1}{|c||}{} &
\multicolumn{3}{c||}{$B_0 = 45$ kG} &
\multicolumn{3}{c||}{$B_0 = 20$ kG} \\
\hline
$\sin 2\theta_0$ & 0.1 & 0.25 & 0.32 & 0.1 & 0.25 & 0.32 \\
\hline
{$\begin{array}{c}{\rm SNO} \\
(\bar{\nu}_e d \rightarrow nne^+) \end{array}$} &
13.1 & 82 & 134 & 7.1 & 44 & 72 \\
\hline
{$\begin{array}{c}{\rm Super-Kamiok.} \\
(\bar{\nu}_e p \rightarrow ne^+ )\end{array}$} &
1380 & 8600 & 14000 & 750 & 4600 & 7600 \\
\hline
{$\begin{array}{c}{\rm Borexino} \\
(\bar{\nu}_e p \rightarrow ne^+ )\end{array}$} &
12 & 74 & 120 & 6.4 & 40 & 65 \\
\hline
\end{tabular}

\vspace{0.5cm}
As few as 5-10 $\bar{\nu}_{eR}$ events/yr in SNO \cite{Robertson} and 20
events/yr in Borexino \cite{RBLBPP} are probably detectable. It therefore
follows from table 1 that for $\sin 2\theta_0 \aprge 0.1$ the solar
$\bar{\nu}_{eR}$'s can be detected in Super--Kamiokande, SNO and Borexino.
The event rates for $B_0=20$ kG are about a
factor of two smaller than those for $B_0=45$ kG. Thus the
$\bar{\nu}_{eR}$ flux should vary in time in $direct$ correlation with
solar activity (11-yr variations). This time dependence of the flux can
facilitate significantly the discrimination between the signal and the
background.

Besides the observable flux of solar $\bar{\nu}_{e}$'s, the predictions
for the forthcoming solar neutrino experiments do not differ much from
those of the RSFP scenario in the absence of neutrino flavor mixing
\cite{Akhm2,ALP1,BG}: strong 11-yr variations of the $^7$Be neutrino flux,
no suppression and no time variations in the neutral--current events
in SNO and moderate time variation ($\aprle 20-30 \%$) of the event rate in
Super--Kamiokande.

In summary, we have shown that the results of the Homestake, Kamiokande and
gallium solar neutrino experiments, including their time structure or lack
of such a structure, can be accounted for provided neutrinos
undergo resonant spin--flavor precession. We have taken possible flavor mixing
of neutrinos into account and demonstrated that the data can only be
reproduced for small enough mixing angles, $\sin 2\theta_0 \aprle 0.25$.
Larger mixing angles are excluded by the data of Ga--Ge experiments. The
solar magnetic field profiles required to fit the data depend substantially
on the neutrino mixing angle. One can discriminate between small mixing angle
($\sin 2\theta_0 \aprle 0.1$) and moderate mixing angle
($0.1\aprle \sin 2\theta_0 \aprle 0.25$) solutions by
studying the solar $\bar{\nu}_{e}$ flux which can be sizeable for moderate
mixing angles. We have calculated the expected $\bar{\nu}_{e}$ signals
in the SNO, Super--Kamiokande and Borexino experiments and showed that they
are detectable even for moderate mixing angles.
\section*{Acknowledgements}
We would like to thank D. Sciama for his stimulating interest in our work.
E.A. and S.P. gratefully acknowledge the kind hospitality of the National
Institute for Nuclear Theory at the University of Washington, where part of
this work has been done. E.A. is grateful to SISSA for warm hospitality and
support. The work of E.A. was supported by the sabbatical grant SAB93-0090
from Spanish Ministry of Education and Science, and that of S.P. was
supported in part by the Bulgarian National Science Foundation via grant
PH-16.

\newpage
\centerline{\bf \large Figure captions}
\vskip 1truecm
\noindent
Fig. 1. The iso-SNU/iso-suppression contours for the Cl--Ar
(Homestake), Kamiokande and Ga--Ge experiments in the
$(B_0,~\Delta m^2)$ plane for $\sin 2\theta_0=0.2$. MAG69+ magnetic field
configuration is used [see eq. (3)]. The full lines are chlorine iso-SNU
curves (1.7, 2.1, 2.55, 3.9, 5.2, 5.7 and 6.4 SNU), the dotted lines
correspond to the ratio of the signal to the reference solar model
prediction \cite{B} for the Kamiokande experiment, $R$=0.30, 0.40, 0.58 and
0.68, and the dash-dotted lines represent the Ga--Ge iso-SNU curves
(53, 66, 79, 92 and 105 SNU). The shaded areas show the allowed ranges of
parameters (see the text).

\vskip 0.5truecm
\noindent
Fig. 2. The iso-SNU/iso-suppression contours in the
($\sin 2\theta_0,~\Delta m^2$) plane for $B_0=20$ kG. The magnetic field
configuration and the definition of the curves are the same as in fig. 1.

\vskip 0.5truecm
\noindent
Fig. 3. Same as in fig.2 but for $B_0=45$ kG.

\vskip 0.5truecm
\noindent
Fig. 4. The iso-SNU/iso-suppression contours in the
($B_0,~\sin 2\theta_0$) plane for $\Delta m^2=10^{-8}$ eV$^2$. The magnetic
field configuration and the definition of the curves are the same as in
figs. 1--3.

\vskip 0.5truecm
\noindent
Fig. 5. Spectra (in MeV$^{-1}$) of $^8$B $\nu_e$'s (full upper curve) and solar
$\bar{\nu}_e$'s (dashed upper curve) normalized to unit integral. The latter
is calculated with MAG69+ magnetic field profile, $\Delta m^2=10^{-8}$
eV$^2$, $\sin 2\theta_0=0.25$ and $B_0=40$ kG. The relative normalization
factor of the two spectra is 36.5. Also shown are the $\nu_{eL}\rightarrow
\bar{\nu}_{eR}$ transition probabilities $P_2$ and $P_{2}^{NVO}$. In the latter
the (anti)neutrino oscillations in vacuum are disregarded (see the text).

\end{document}